\documentclass[12pt]{iopart}
\usepackage{cite}
\usepackage{amsfonts}
\usepackage{amssymb}
\usepackage{graphicx}
\usepackage{bbold}
\usepackage{textcomp}
\usepackage{color}
\usepackage{xfrac}
\usepackage{subfigure}
\newcommand{\PrYSO}{Pr$^{3+}$:Y$_2$Si{O$_5$}}
\newcommand{\EuYSO}{Eu$^{3+}$:Y$_2$Si{O$_5$}}
\newcommand{\gguide}{{\it Coherent Storage of Multimode Light Using a Spin-Wave AFC Memory}}
\newcommand{\mysize}{\scriptsize}

\usepackage{iopams}
\begin{document}
\graphicspath{{figuresPaper/}{}}
\title[\gguide]{Coherent Storage of Temporally Multimode Light Using a Spin-Wave Atomic Frequency Comb Memory}
\author{M.~G\"{u}ndo\u{g}an$^{1}$, M.~Mazzera$^{1}$, P.~M.~Ledingham$^{1}$, M.~Cristiani$^{1}$,  H.~de~Riedmatten$^{1,2}$}
\address{$^{1}$ICFO-Institut de Ciencies Fotoniques, Av. Carl Friedrich Gauss 3, 08860 Castelldefels (Barcelona), Spain.}
\address{$^{2}$ICREA-Instituci\'{o} Catalana de Recerca i Estudis Avan\c{c}ats, 08015 Barcelona, Spain} 
\ead{patrick.ledingham@icfo.es}
\begin{abstract}
We report on coherent and multi-temporal mode storage of light using the full atomic frequency comb memory scheme. The scheme involves the transfer of optical atomic excitations in {\PrYSO} to spin-waves in the hyperfine levels using strong single-frequency  transfer pulses. Using this scheme, a total of {5} temporal modes are stored and recalled on-demand from the memory. The coherence of the storage and retrieval is characterized using a time-bin interference measurement resulting in visibilities  higher than 80$\%$, independent of the storage time. This coherent and multimode spin-wave memory is promising as a quantum memory for light.
\end{abstract}

\section{Introduction}
The coherent, efficient and reversible mapping between quantum light and matter represents a fundamental challenge in the field of  quantum information science. Overcoming this challenge would enable the realization of a quantum memory for light, a device that can store and, on-demand, recall quantum states of light with high efficiency and fidelity.  Such memories have potential applications for use in a quantum repeater \cite{Briegel1998, Duan2001,Sangouard2011}, a resource allowing for quantum communication over long distances. Other applications include linear optics quantum computation \cite{Kok2007}, deterministic single photon sources \cite{Matsukevich2006} and multi-photon quantum state engineering \cite{Nunn2012}. For practical applications, it is likely that a quantum memory with a high multimode capacity will be necessary.

An example of a quantum memory for light is a single atom trapped in a cavity \cite{Specht2011}. An alternate candidate is the use of atomic ensembles \cite{Simon2010}. The benefit of using such systems is that the light can be absorbed collectively, enhancing the coupling between light and matter. Indeed, extensive research has been put into single atom and atomic ensemble based memories alike. Realizations of ensemble based memories include cold \cite{Chaneliere2005, Chou2005,Simon2007c,Choi2008,Radnaev2010,Bao2012} and hot atomic gases \cite{Julsgaard2004,Eisaman2005, Reim2011,Hosseini2011}, and solid state systems \cite{Riedmatten2008, Hedges2010,Amari2010,Sabooni2010,Chaneliere2010,Lauritzen2010,Usmani2010,Bonarota2011, Clausen2011,Saglamyurek2011,Usmani2012,Saglamyurek2012,Gundogan12, Clausen2012, Zhou2012,Ledingham2012,Beavan2012,McAuslan2011,Damon2011}.

A promising ensemble based quantum memory for light is the atomic frequency comb memory (AFC) \cite{Afzelius2009} based on inhomogeneously broadened media, such as cryogenically cooled rare earth ion doped crystals. The AFC memory requires fine spectral tailoring of the inhomogeneously broadened absorption line into a series of equally spaced, narrow absorbing peaks.  A resonant input pulse whose bandwidth matches that of the comb is collectively absorbed. This atomic coherence initially dephases, but due to the periodic structure of the comb, rephases and coherently re-emits an echo. This light is in the same spatial mode as the input and at a delayed time of $\tau = \sfrac{1}{\Delta}$, where $\Delta$ is the spectral distance between the peaks.  

Recent progress towards solid state quantum memories using the AFC scheme include the storage of weak coherent pulses at the single photon level \cite{Riedmatten2008,Sabooni2010,Chaneliere2010,Lauritzen2010} and the storage of multiple temporal modes in a crystal \cite{Usmani2010,Bonarota2011}. Storage of quantum light has also been demonstrated. The use of nonclassical light generated by spontaneous down conversion has enabled entanglement between one photon and one collective atomic excitation stored in a crystal \cite{Clausen2011,Saglamyurek2011}, entanglement between two crystals \cite{Usmani2012}, and time-bin qubit storage \cite{Saglamyurek2012}. Recently, the versatility of these memories was extended to the quantum storage of polarization qubits \cite{Gundogan12, Clausen2012, Zhou2012}. 

\subsection{Three Level AFC Scheme}
All of the above realizations used AFC storage in the excited state. However, strictly speaking, it is not an on-demand memory for light but rather a pre-programmed delay. In addition, the delay is limited to the excited state lifetime. On-demand retrieval of light, which is necessary for applications in quantum information science, can be achieved with a full AFC scheme which, as proposed in \cite{Afzelius2009}, transfers coherently the optical atomic excitation to and from a spin excitation. This proposal requires the use of three ground states. One is for the initial state (comb), one for the spin-wave excitations and a third one is needed as an auxiliary state for unnecessary ions that are not part of the AFC. The coherence needs to be transferred before the re-emission time $\tau$. Once the coherence is transferred to a spin-wave, for example by a  single frequency $\pi$-pulse, the dipole evolution is effectively frozen. To read out this spin-wave, a second transfer pulse is applied after a time $T_S$, resulting in a three level echo (3LE) with a total storage time of $\tau + T_S$. If each transfer pulse has an efficiency of $\eta_{\textrm{T}}$, then the total efficiency is given by \cite{Afzelius2010}
\begin{equation}
\eta_{\textrm{3LE}} = \eta_{\textrm{AFC}} \, \eta_{\textrm{T}}^2,
\label{eq:3LEeff}
\end{equation}
where $\eta_{\textrm{AFC}}$ is the 2 level AFC echo efficiency. Assuming Gaussian absorbing peaks, the efficiency in the forward direction is approximated well by 

\begin{equation}
\eta_{\textrm{AFC}} \approx {\tilde{d}}^2 \, e^{-\sfrac{7}{F^2}} \, e^{-\tilde{d}}\, e^{-d_0}, 
\end{equation}

where $F = \sfrac{\Delta}{\gamma}$ is the finesse of the comb with $\gamma$ being the peak width,  $\tilde{d} = \sfrac{d}{F}$ is the effective optical depth experienced by the absorbed pulse with $d$ being the optical depth and $d_0$ is the absorbing background \cite{Afzelius2009}. The transfer efficiency can be optimized by use of chirped pulses as theoretically discussed in \cite{Minar2010}.

An additional benefit  of the full AFC scheme is given by the presence of an extra degree of freedom related to the propagation direction of the control pulse. This allows for backward retrieval of the echo, enabling in theory a storage and retrieval efficiency of $100\,\%$ \cite{Afzelius2009}. Also, spatial separation of the control mode from the input mode is available. 

A proof of principle of the full AFC scheme has been demonstrated using praseodymium doped yttrium oxyorthosilicate ({\PrYSO}) with a total efficiency of $0.5 - 1\,\%$ (transfer efficiency of  $30 - 45\,\%$), with a multimode capacity of 2 temporal modes \cite{Afzelius2010}. More recently, the full AFC scheme has been implemented in a different rare earth ion sample ({\EuYSO}) \cite{Timoney2012} with a lower efficiency due to the low oscillator strength of the {Eu$^{3+}$} transition compared to {Pr$^{3+}$}. In this paper, using {\PrYSO}, we confirm for the first time the coherent nature of the storage explicitly by way of a time-bin interference measurement. Furthermore, we extend the efficiency and multimode capacity beyond what has been demonstrated previously. For this demonstration we use bright pulses, however the scheme is in principle extendable to the use of single photons as we will discuss later.

\subsection{Praseodymium}

Cryogenically cooled {\PrYSO} is an attractive medium for solid state quantum memories. As all the rare earth ions,  {Pr$^{3+}$} is characterized by a partially filled 4f shell spatially located within the full 5s and 5p ones. This feature allows {Pr$^{3+}$} to maintain an atomic-like energy level scheme, with homogeneous linewidth on the order of $2\,\mathrm{kHz}$ \cite{Equall1995, Sun2005}, even when embedded in a crystalline matrix. Furthermore, the low nuclear magnetic moment of the neighbouring ions (Si and Y) prevents spin-flips and magnetic ion-host interactions. A second order hyperfine interaction is able to split the crystal field singlets into three sub-levels, providing the three-fold ground state required by the full AFC scheme. The optical transition used for the storage protocol  connects the lowest-lying levels (labelled with (1)) of the ground  $^3\textrm{H}_4$ and excited ${^1}\textrm{D}_2$ manifolds  (see Fig. \ref{fig1b}), characterized by a wavelength of $605.977\, \mathrm{nm}$. In samples doped with $0.02\,\%$  {Pr$^{3+}$}, it has been reported to exhibit an inhomogeneous bandwidth of $\sim\,5\,\mathrm{GHz}$, which allows for the tailoring of the absorption profile into a frequency comb structure. The excited state has a lifetime of $T_1 = 164\, \mu\mathrm{s}$ and the coherence time with zero magnetic field is $T_2 = 111\,\mu\mathrm{s}$ \cite{Equall1995}. With the use of specific external magnetic fields \cite{Fraval2004} and dynamical decoupling sequences \cite{Souza2011}, light has been stored using electromagnetically induced transparency  in {\PrYSO} for times exceeding $1\,\mathrm{s}$ \cite{longdell2005} and a hyperfine ground state coherence time exceeding $30\,\mathrm{s}$ has been demonstrated using similar techniques \cite{Fraval2005}.

\section{Experimental Description}
\subsection{Optical Set Up}
\begin{figure}[h!]
\centering
\subfigure[\label{fig1a}]{
\def\svgwidth{0.8\textwidth}
\begingroup
  \makeatletter
  \providecommand\color[2][]{
    \renewcommand\color[2][]{}
  }
  \providecommand\transparent[1]{
   \renewcommand\transparent[1]{}
  }
  \providecommand\rotatebox[2]{#2}
  \ifx\svgwidth\undefined
    \setlength{\unitlength}{357.30991211pt}
    \ifx\svgscale\undefined
      \relax
    \else
      \setlength{\unitlength}{\unitlength * \real{\svgscale}}
    \fi
  \else
    \setlength{\unitlength}{\svgwidth}
  \fi
  \global\let\svgwidth\undefined
  \global\let\svgscale\undefined
  \makeatother
  \begin{picture}(1,0.21452689)
    \put(0,0){\includegraphics[width=\unitlength]{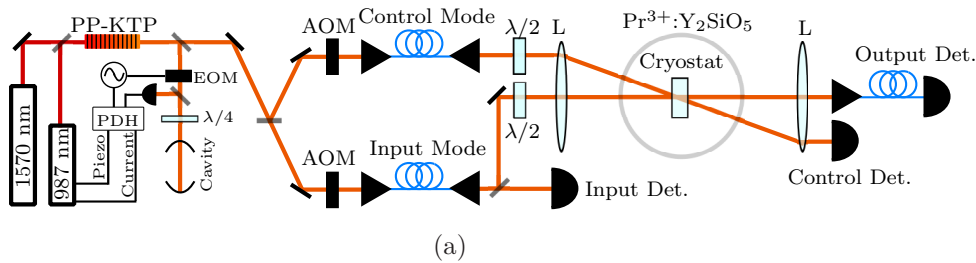}}
    \put(0.37,0.2){\color[rgb]{0,0,0}\makebox(0,0)[lb]{\smash{\mysize Control Mode}}}
    \put(0.38175702,0.06078502){\color[rgb]{0,0,0}\makebox(0,0)[lb]{\smash{\mysize Input Mode}}}
    \put(0.6675,0.15303741){\color[rgb]{0,0,0}\makebox(0,0)[lb]{\smash{\mysize Cryostat}}}
    \put(0.65,0.195){\color[rgb]{0,0,0}\makebox(0,0)[lb]{\smash{\mysize \PrYSO}}}
    \put(0.83573721,0.18782317){\color[rgb]{0,0,0}\makebox(0,0)[lb]{\smash{\mysize L}}}
    \put(0.575,0.18846757){\color[rgb]{0,0,0}\makebox(0,0)[lb]{\smash{\mysize L}}}
    \put(0.82490348,0.02984834){\color[rgb]{0,0,0}\makebox(0,0)[lb]{\smash{\mysize Control Det.}}}
    \put(0.89589088,0.1597913){\color[rgb]{0,0,0}\makebox(0,0)[lb]{\smash{\mysize Output Det.}}}
    \put(0.610336,0.01681974){\color[rgb]{0,0,0}\makebox(0,0)[lb]{\smash{\mysize Input Det.}}}
    \put(0.525,0.18866756){\color[rgb]{0,0,0}\makebox(0,0)[lb]{\smash{\mysize $\lambda$/2}}}
    \put(0.525,0.07820441){\color[rgb]{0,0,0}\makebox(0,0)[lb]{\smash{\mysize $\lambda$/2}}}
    \put(0.022,0.02509299){\color[rgb]{0,0,0}\rotatebox{90}{\makebox(0,0)[lb]{\smash{\mysize 1570 nm}}}}
    \put(0.06308496,0.01306122){\color[rgb]{0,0,0}\rotatebox{90}{\makebox(0,0)[lb]{\smash{\mysize 987 nm}}}}
    \put(0.31, 0.05){\color[rgb]{0,0,0}\makebox(0,0)[lb]{\smash{\mysize AOM}}}
    \put(0.31, 0.19){\color[rgb]{0,0,0}\makebox(0,0)[lb]{\smash{\mysize AOM}}}
    \put(0.065,0.1908224){\color[rgb]{0,0,0}\makebox(0,0)[lb]{\smash{\mysize PP-KTP}}}
    \put(0.205,0.095){\color[rgb]{0,0,0}\makebox(0,0)[lb]{\smash{\tiny $\lambda$/4}}}
    \put(0.195,0.137){\color[rgb]{0,0,0}\makebox(0,0)[lb]{\smash{\tiny EOM}}}
    \put(0.215,0.02086062){\color[rgb]{0,0,0}\rotatebox{90}{\makebox(0,0)[lb]{\smash{\tiny Cavity}}}}
    \put(0.10134592,0.03403319){\color[rgb]{0,0,0}\rotatebox{90}{\makebox(0,0)[lb]{\smash{\tiny Piezo}}}}
    \put(0.13145798,0.015){\color[rgb]{0,0,0}\rotatebox{90}{\makebox(0,0)[lb]{\smash{\tiny Current}}}}
    \put(0.093,0.0915){\color[rgb]{0,0,0}\makebox(0,0)[lb]{\smash{\tiny PDH}}}
  \end{picture}
\endgroup
 }
 \subfigure[\label{fig1b}]{
\includegraphics[width=.75\textwidth]{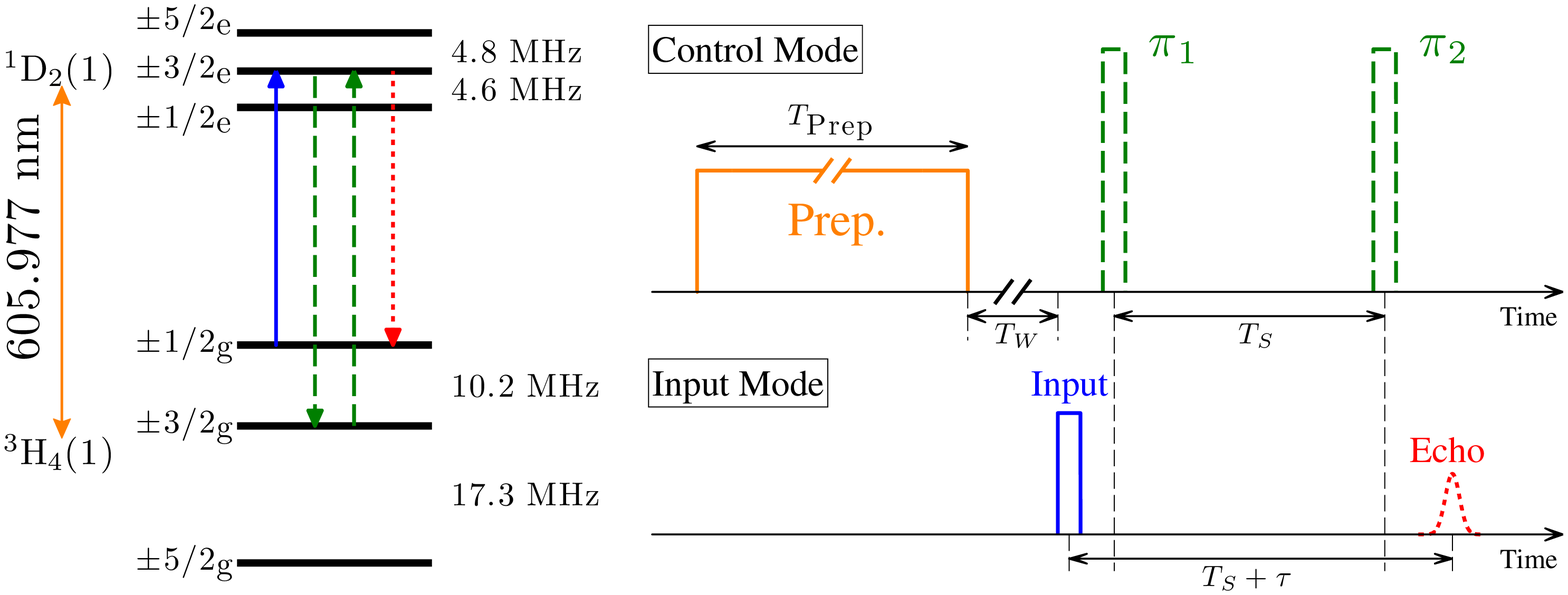}
    }
\caption{ (a) Experimental Setup. See text for details. $\lambda/2 \,(\lambda/4)$: Half (quarter) wave plate, L: $100\,\mathrm{mm}$ lens, AOM:  acousto-optic modulator in double pass configuration, EOM: electro-optic modulator driven at $12.5\,\mathrm{MHz}$, PDH: Pound-Drever-Hall module for laser frequency locking. (b) Level scheme and pulse sequence for the spin-wave AFC memory. The comb is prepared in a time $T_{\textrm{Prep}} = 200\,\mathrm{ms}$. A waiting time of $T_W = 1\,\mathrm{ms}$ is used between the end of the preparation and the start of the 3LE sequence. $T_s$ is the time between the control pulses $\pi_1$ and $\pi_2$, and $\tau$ is the AFC delay time. The total storage time is $T_S + \tau$.}
\end{figure}

Figure \ref{fig1a} shows the experimental set up. Our laser source is based on sum frequency generation of $987\,\mathrm{nm}$  and $1570\,\mathrm{nm}$ light. These wavelengths are combined in a periodically poled (PP) KTP waveguide. With input power of $429\,\mathrm{mW}$ and $1138\,\mathrm{mW}$ for the $987\,\mathrm{nm}$ laser and $1570\,\mathrm{nm}$ laser respectively, we achieve an output power of $190\,\mathrm{mW}$ at $606\,\mathrm{nm}$, corresponding to a conversion efficiency of $389\,\%/\mathrm{W}$.

Acousto-optic modulators (AOM) are used in double pass configuration to create the necessary pulsed light for the echo experiments. This configuration allows to scan the frequency of the light by several MHz while keeping the intensity flat and spatial mode the same. The RF signal used to drive the AOMs are generated using an arbitrary waveform generator (Signadyne). We use two different optical paths to interact with the sample, the control and input modes, each having a double pass AOM. The advantage of this choice is two-fold. On one hand, having a mode for the preparation and strong control light spatially separated from the input and weak echo light, prevents noise from the strong control polluting the echo mode. On the other hand, a strong control pulse might cause a free induction decay (FID) due to off resonant excitation or an incorrectly prepared transparency window (see subsection \ref{ssec:memprep}). The echo would be hidden by this additional noise if the two modes were not spatially separated. These advantages become important in perspective of single photon level inputs.

After each AOM, the light is coupled to a single mode fiber and then out-coupled on an isolated optical bench, where the closed cycle cryogenic cooler (Oxford V14) used to cool the crystal  to $2.8\,~\mathrm{K}$ is located. Our {\PrYSO} sample is $3\,\mathrm{mm}$ thick with 0.05$\,\%$ doping. The absorption coefficient is measured to be $\alpha = 23\,\mathrm{ cm}^{-1}$ for the optical transition at $605.977\,$nm. The inhomogeneous linewidth is measured to be $5\,\mathrm{GHz}$. Half-wave plates ($\lambda/2$) ensure the polarization is aligned parallel to the optical D$_2$ axis of the crystal, in order to maximize the absorption \cite{Sun2005}. A glass plate is placed in the input mode allowing a reference signal on the input detector. An $f = 100\,\mathrm{mm}$ lens is used to focus both modes on to the crystal, resulting in a beam width size of  around {$95\,\mu\mathrm{m}$} for both the control  and input modes. The maximum power in the control mode before the cryostat window is measured to be $7\,\mathrm{mW}$. The maximum power available in the input mode is $1.5\,\mathrm{mW}$. An additional $f = 100\,\mathrm{mm}$ lens is placed after the crystal for spatial mode matching. The echo mode is steered to a detector via a single mode fiber with {$50\,\%$} coupling efficiency.

{The $606\,\mathrm{nm}$ light is frequency locked to a passively stabilized cavity using the Pound-Drever-Hall locking technique \cite{Drever1983}. This method of locking  provides the desirable laser linewidth for the fine spectral tailoring of the AFC, as well as long term frequency stability.} The set up schematic is shown in Fig. \ref{fig1a}. The cavity has a free spectral range of $1\,\mathrm{GHz}$ and a linewidth of $500\,\mathrm{kHz}$  leading to a finesse of $2000$. A temperature stabilized Invar spacer is used to hold the cavity mirrors. Light used for locking is picked off before the AOMs and directed through an electro-optic modulator (EOM). The EOM is driven by an RF source (Toptica Digilock module) resulting in phase-modulated sidebands at $12.5\,\mathrm{MHz}$. The error signal is created using in-house electronics and is fed-back to the piezo and current drivers of the 987 nm laser via the Toptica Digilock and FALC modules respectively. Spectral hole-burning experiments reveal the laser linewidth to be around $100\,\mathrm{kHz}$ for ms timescales.

\subsection{Memory Preparation\label{ssec:memprep}}

The scheme of the full AFC protocol realized in the present work is depicted in Figure \ref{fig1b}. The input pulse (solid arrow) is resonant with the comb on the ${1}/{2}_\textrm{g}-{3}/{2}_\textrm{e}$ transition, while the coherent transfer of the optical excitation to and from the ground state (dashed arrows) is tuned to the ${3}/{2}_\textrm{g}-{3}/{2}_\textrm{e}$  transition. The echo is emitted on the ${1}/{2}_\textrm{g}-{3}/{2}_\textrm{e}$ transition (dotted arrow). The remaining ${5}/{2}_\textrm{g}$ ground level is exploited as the auxiliary state. 

The preparation of the memory follows the approach described in ref. \cite{Nilsson2004}. A transparency window is first created inside the absorption profile by sweeping the laser frequency by $12\,\mathrm{MHz}$ using an AOM, thus pumping all the ions which are resonant with this light to a non-resonant ground state. The sweep is repeated {100} times using the strong control mode, thus creating a power broadened $18\,\mathrm{MHz}$ wide transparency window (the ``pit''). To tailor the AFC, the burn-back method is used. It consists of the pumping of ions from the $5/2_\textrm{g}$ auxiliary state, some of which will decay in the ground states previously emptied giving rise to isolated narrow peaks in the transparency window. By sending several burn-back pulses at frequencies differing by $\Delta$, the AFC structure is created. The finesse of the comb, and thus the efficiency of the AFC echo, can be optimized with a proper choice of duration, power and number of the burn-back pulses. 

It is important to note that, since the hyperfine level splittings ($\sim$\,MHz) are smaller than the inhomogeneous width ($\sim$\,GHz), different classes of ions will be resonant with the burn-back pulses \cite{Guillot-Noel2007}. In the frequency window of interest for the comb, transitions associated to two classes of ions are indeed present (namely, ~${1}/{2}_\textrm{g}-{3}/{2}_\textrm{e}$  of class I and  ${1}/{2}_\textrm{g}-{1}/{2}_\textrm{e}$  of class II).

The burn-back procedure also populates the ${3}/{2}_\textrm{g}$ ground state, while an efficient full AFC scheme requires it to be empty. As a matter of fact, if some ions remained in the ground state addressed by the control pulse, they would be driven to the excited state and give noise in the echo mode due to spontaneous emission. To avoid this additional noise, a further ``clean'' sweep is performed in the spectral region of the ${3}/{2}_\textrm{g}-{3}/{2}_\textrm{e}$  transition whose bandwidth matches that of the comb. This clean sweep offers also the ulterior advantage of emptying the ground states of the unwanted classes.

In this paper, the total preparation time, $T_{\textrm{Prep}}$ from Fig. \ref{fig1b}, is $200\,\mathrm{ms}$, of which $50\,\mathrm{ms}$ are used to create the comb.  {This includes $5\times100\,\mu\mathrm{s}$ burn back pulses for the 5 comb peaks repeated every $500\,\mathrm{ns}$ for $100$ times and a $2\,\mathrm{MHz}$ clean sweep centered at the transfer pulse frequency repeated 1000 times. 

For the experimental results presented in the next section, we use different detection methods. For the results presented in subsections \ref{ssec:AFC}-\ref{ssec:coherent} a photodiode (Thorlabs PDB150) is used and a memory is prepared for each pulse that is stored. For subsection \ref{ssec:MM} a single photon counter (Laser Components, model Count) is used and for each memory prepared, 500 pulse trains are used.

\begin{figure}[h!]
\begin{center}
\subfigure[\label{fig2a}]{\includegraphics[width=0.45\textwidth]{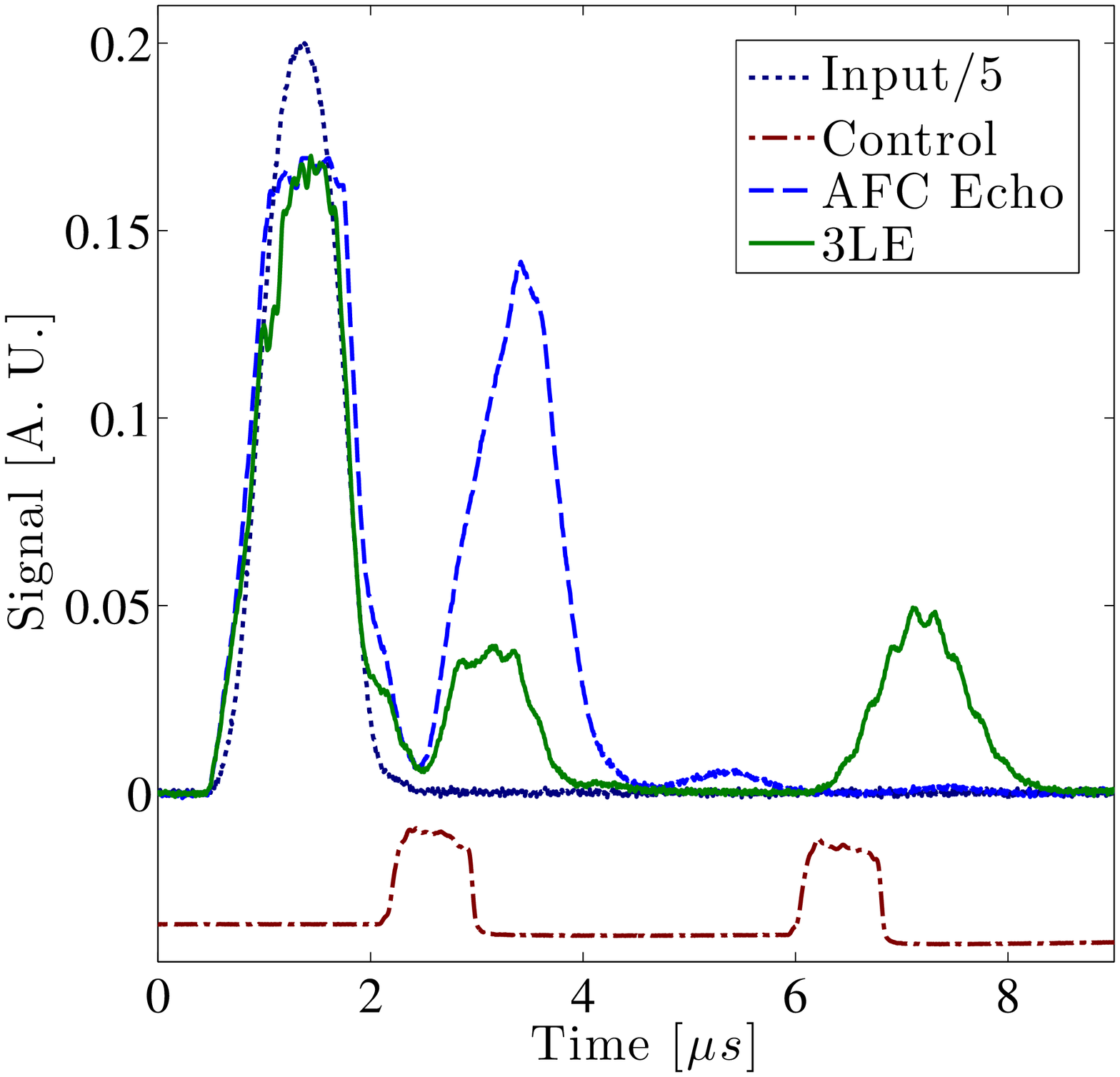}}
\subfigure[\label{fig2b}]{\includegraphics[width=0.44\textwidth]{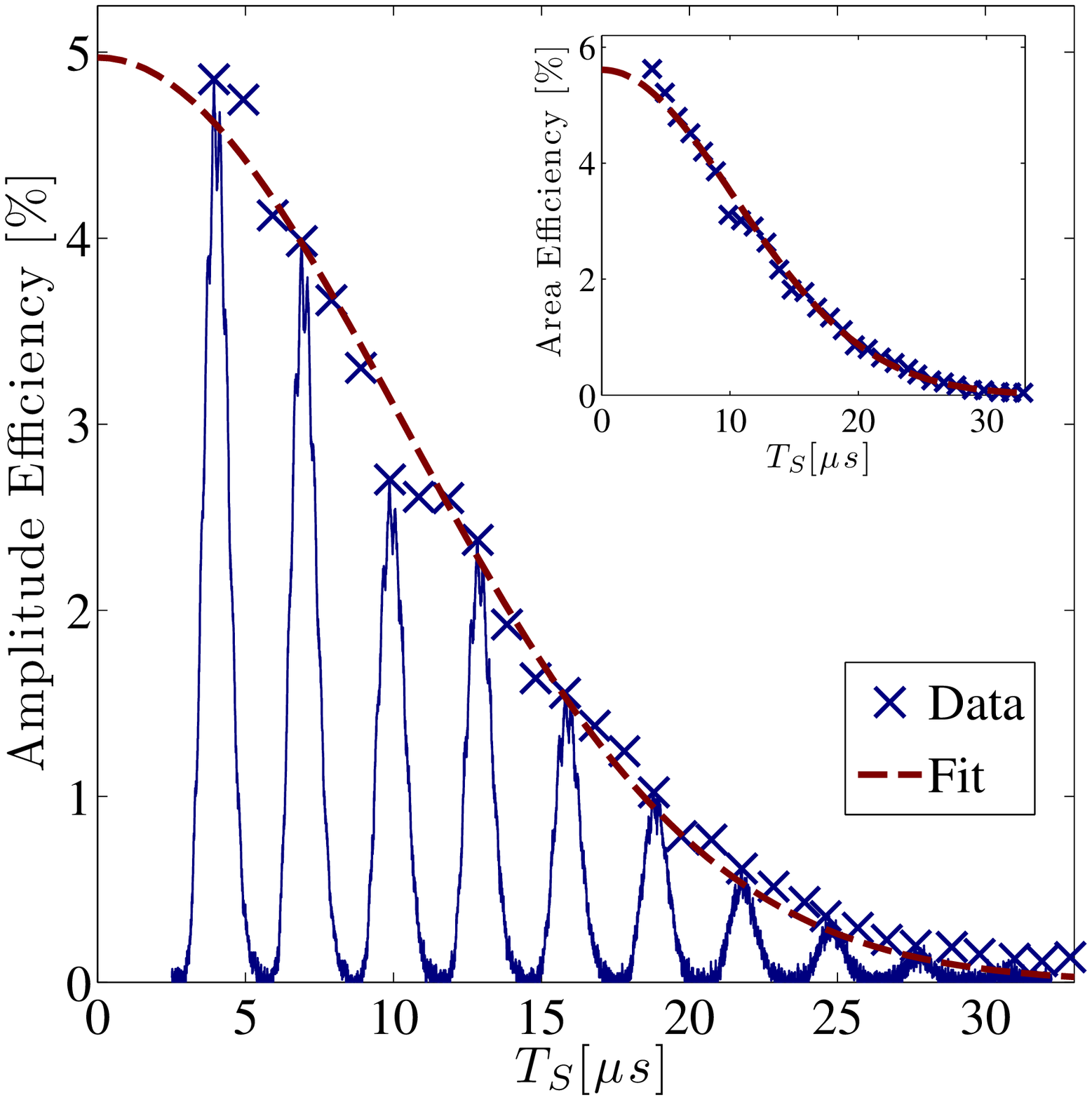}}
\caption{(a)The full AFC memory scheme. The blue dashed curve shows the two level echo from an AFC programmed to give a delay of $2\, \mu\mathrm{s}$ $(\Delta = 500\,\mathrm{kHz})$. The green solid curve shows the effect of the transfer pulses. The AFC echo is reduced and the 3LE is observed. The dark blue dotted trace shows a reference of the input pulse, with full width at half max duration of  {$840\,\mathrm{ns}$}. For clarity, this curve is divided by 5. The dot-dashed trace indicates the temporal location of the {$800\,\mathrm{ns}$} rephasing pulses separated by $T_S = 4\,\mu\mathrm{s}$. (b)  3LE efficiency vs. $T_S$ for the same conditions as in Fig. \ref{fig2a}. All data are an average of 10 experimental runs.}
\label{fig2}
\end{center}
\end{figure}

\section{Results}

\subsection{AFC Storage\label{ssec:AFC}}

Figure \ref{fig2a} shows an example of a two-level AFC echo (dashed curve) for an excited state storage time of $\tau = 2\, \mu\mathrm{s}$. Using input pulses of $840\,\mathrm{ns}$ duration, an AFC efficiency of {$\eta_{\mathrm{AFC}} = 15.6 \pm 0.5\,\%$} is observed. The efficiency is measured by taking the area of the echo in the blue dashed trace and comparing this to the area of the input pulse when sent through a spectral pit (dark blue dotted trace). When measuring the input pulse area, its polarization is rotated to be perpendicular to the optical D$_2$ axis, ensuring weak interaction with any residual ions in the pit. To estimate the parameters of the comb, we compare the area of the transmitted pulse to that of the echo \cite{Riedmatten2008}. Assuming a tooth width of $\gamma =125 \, (167)\,\mathrm{kHz}$, then $F = 4 \, (3)$ and the inferred optical depths are $d = 4.12\, (3.66)$ and $d_0 = 0.45 \, (0.26)$.

\subsection{Spin-Wave Storage\label{ssec:3LE}}

Figure \ref{fig2a} shows an example of the full AFC scheme of spin-wave storage (solid curve). A strong single frequency control pulse is used to transfer the excited state coherence to and from the ${3}/{2}_\textrm{g}$ ground state. The effect of the first transfer pulse is clearly seen by the significant reduction of the AFC echo. A second transfer pulse is applied after a time $T_S = 4\,\mu\mathrm{s}$ and a 3LE is clearly seen above the noise. It is the direct result of the application of both control pulses, as confirmed by the fact that it vanishes by removing either $\pi_1$, $\pi_2$ or the input itself. 

Figure \ref{fig2b} shows the 3LE efficiency, both amplitude and area (inset), as a function of $T_S$. Due to the inhomogeneous broadening of the spin state, the efficiency of the echo decays as a function of $T_S$. This inhomogeneity is known to be Gaussian with the following form \cite{Timoney2012}:
\begin{equation}
\eta(T_S)_{\textrm{3LE}} = \eta(0)_{\textrm{3LE}}\times\exp\left[{\frac{-(\gamma_{\textrm{IS}}  T_S)^2}{{2 \log(2)}/{\pi^2}}}\right],\label{eqIS}
\end{equation}
where $\gamma_{\textrm{IS}}$ is the spin inhomogeneous broadening and $\eta(0)_{\textrm{3LE}}$ is the 3LE efficiency at zero delay. Fitting the data with this equation and extrapolating to $T_S = 0$ where the spin inhomogeneity has no effect, the 3LE efficiency ($\eta(0)_{\textrm{3LE}}$) for the area (amplitude) is $5.6 \pm 0.1\,\%\, (5.0 \pm 0.1\,\%)$. The corresponding inhomogeneity is $\gamma_{\textrm{IS}} = 25.6 \pm 0.2\,\mathrm{kHz}$ $(25.7 \pm 0.4\,\mathrm{kHz})$ in agreement with previous realizations \cite{Ham2003, Afzelius2010}. Comparing the AFC efficiency with this 3LE efficiency, the transfer efficiency is calculated to be $57\,\% - 60\,\%$ using Eq. \ref{eq:3LEeff}. 

An alternate method to assess the transfer efficiency is to compare the area of the AFC echo to the one reduced by the first control pulse. From Fig. \ref{fig2a} we get $\eta_T = 73 \, \%$ for $T_S = 4\,\mu\mathrm{s}$.  This is higher than what is obtained from the 3LE. The disagreement could be due to the imperfect spatial mode overlap between the control and input modes.

This is the full AFC scheme with the highest efficiency reported so far. Nevertheless, in view of applications in quantum communication, further improvements are necessary as we discuss later.

\subsection{Transfer Characterization\label{ssec:transfer}}

We now characterize the transfer efficiency $\eta_T$ as a function of the power in the control pulse and thus Rabi frequency ($\Omega_R$). The experimental conditions for this measurement are identical to those outlined in Fig. \ref{fig2a}. A time $T_S$ of $4\,\mu\mathrm{s}$ is chosen as a time fast compared to the spin inhomogeneous dephasing time so as not to limit the efficiency.  
\begin{figure}[h!]
\begin{center}
{\includegraphics[width=0.75\textwidth]{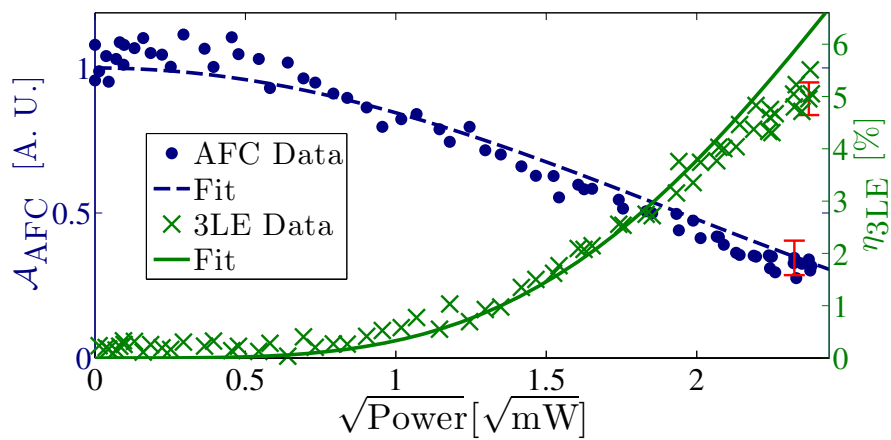}}
\caption{Two level AFC echo area $\mathcal{A}_{\textrm{AFC}}$ (dots) and three level efficiency $\eta_{\textrm{3LE}}$ (crosses) as a function of the square root of power of the control pulses. The dashed and solid lines represent a fit to the data as described in the text. Each data point is an average of 10 trials. Error bars indicate the maximum standard error in the mean measured for each data set. }
\label{fig3}
\end{center}
\end{figure}

Figure \ref{fig3} shows the 3LE efficiency and the area of the AFC echo $\mathcal{A}_{\textrm{AFC}}$ as a function of the square root of power of the control pulse.  Intuitively, as the power is increased the AFC echo decreases and the 3LE increases. This behaviour can be modeled simply with a two level atomic system being driven resonantly \cite{Allen1975, Scully1997}, resulting in the following equations
\begin{eqnarray}
\mathcal{A}_{\textrm{AFC}}(\bar{\Omega}_R t) &= \mathcal{A}_{\textrm{IN}}\frac{\eta_{\textrm{AFC}}}{2}\,\left[ 1 + \cos(\bar{\Omega}_R t)\right],\label{eqRabi2LE}\\
\eta_{\textrm{3LE}}(\bar{\Omega}_R t) &= \frac{\eta_{\textrm{AFC}}}{4}\,\left[ 1 - \cos(\bar{\Omega}_R t)\right]^2,\label{eqRabi3LE}
\end{eqnarray}
where $\mathcal{A}_{\textrm{IN}}$ is the area of the input pulse and we introduce $\bar{\Omega}_R$ as the effective Rabi frequency. This takes into account that the input and control pulses have inhomogeneous spectral and spatial profiles and also the imperfect spatial mode overlap between the two modes. The data in Fig. \ref{fig3} are fit simultaneously with the above equations and show good agreement. The effective Rabi frequency is found to be {$\bar{\Omega}_R~=~2\pi\times340\,\mathrm{kHz}$} for a power of $5.7\,\mathrm{mW}$ at the crystal. 

As a comparison, we measure the Rabi frequency with coherent hole burning techniques \cite{Lauritzen2012} using only the control mode, resulting in $2\pi \times 420\,\mathrm{kHz}$. Finally, knowing the oscillator strength of the transition  \cite{Nilsson2004, Guillot-Noel2007} and the laser intensity, we calculate the expected Rabi frequency to be about $2\pi \times 430\,\mathrm{kHz}$. This Rabi frequency is independent of spectral inhomogeneity and spatial mode overlap, therefore is higher than the effective Rabi frequency. 
 
\subsection{Coherent Storage\label{ssec:coherent}}

In order to implement a quantum memory, it is crucial that the phase between 2 modes is preserved during storage and retrieval. For quantum communication schemes, information is often encoded into the amplitude and phase of time-bin qubits, as they are known to be robust against decoherence in optical fibers \cite{Gisin2002}. We confirm for the first time the coherent storage of time-bin information using the spin-wave AFC memory.
\begin{figure}[h!]
\begin{center}
\subfigure[\label{fig4a}]{\includegraphics[width=0.36\textwidth]{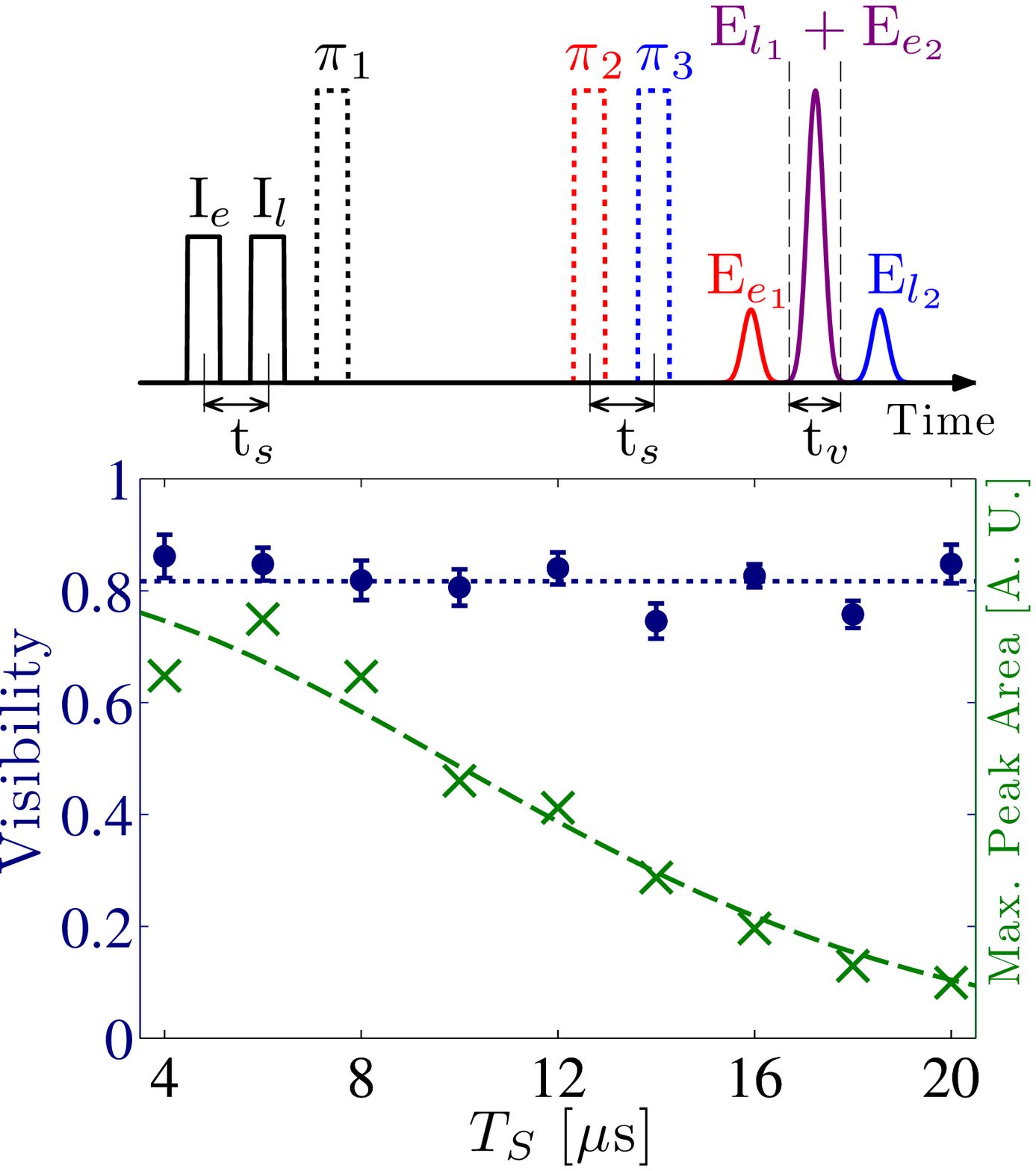}}
\subfigure[\label{fig4b}]{\includegraphics[width=0.56\textwidth]{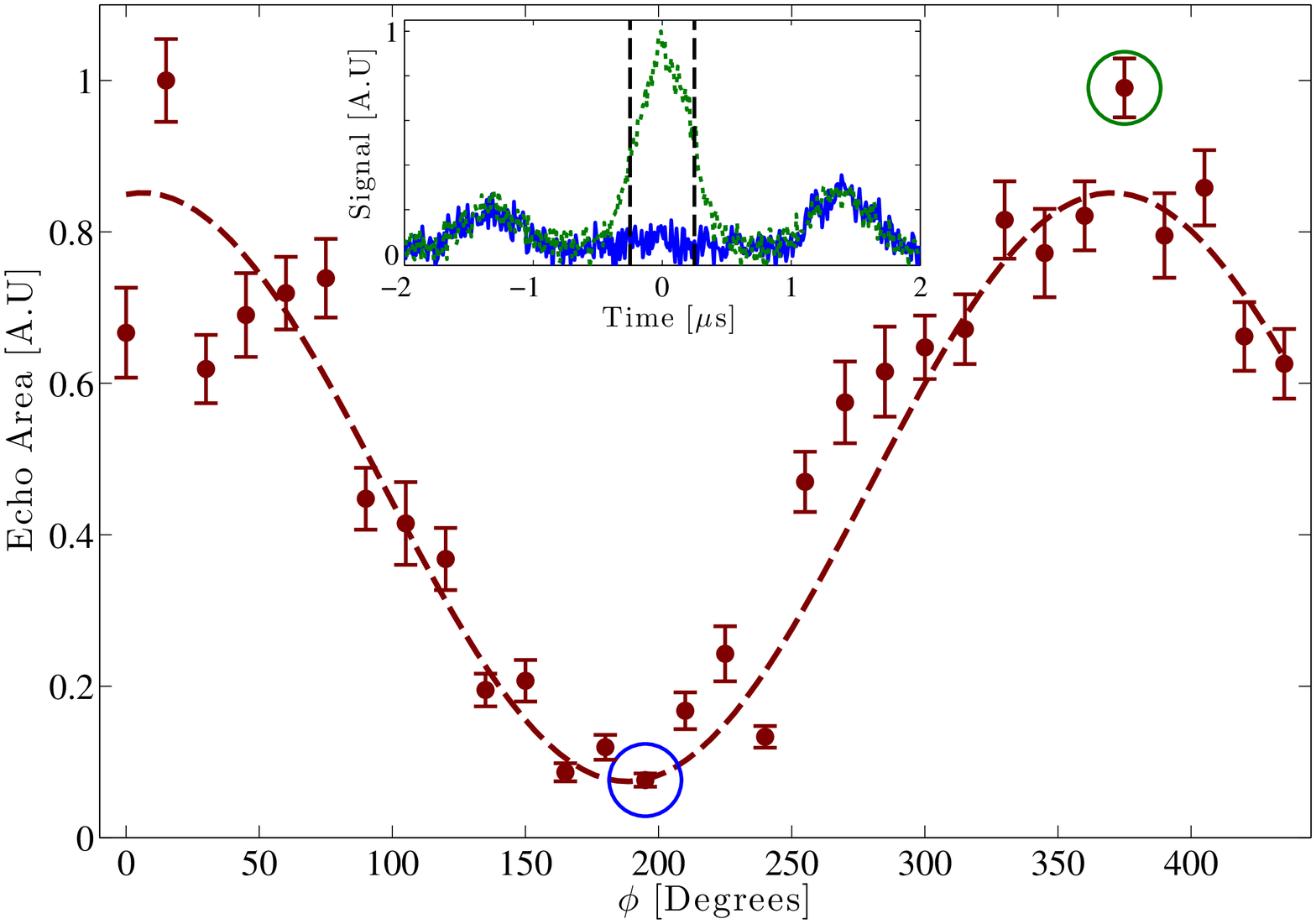}}
\caption{(a) Upper panel: Pulse sequence used for the interference experiment. Lower panel: Visibility and maximum peak area vs. $T_S$. The dots (crosses) correspond to the measured visibility (area). For each $T_S$, a curve like (b) is measured from which the visibility is extracted via a weighted fit. The error bar is a confidence interval of this fit of one standard deviation. The horizontal dotted line is the mean of the visibilites, $\bar{\textrm{V}}=81.7 \pm 1.4\,\%$. The dashed line is a fit using Eq.~\ref{eqIS} with a corresponding spin inhomogeneity of $\gamma_{\textrm{IS}} = 26.8 \pm 1.7\,\mathrm{kHz}$. (b) Echo area in the interference window vs. phase difference between the two input pulses. Each data point is an average over 10 trials with the error bar representing one standard deviation of the mean. The dashed line is a weighted fit of the data to a sine. The visibility is $84.0 \pm 5.7\,\%$. Inset: Time-bin qubit interference for a phase difference of $195^\circ$ (solid line) and $375^\circ$ (dotted line). The corresponding data points in the main figure are labelled with circles. The vertical dashed lines indicate the interference window (t$_v = 500\,\mathrm{ns}$).}
\label{fig5}
\end{center}
\end{figure}

To verify the coherence of the storage, we send a pair of pulses (I$_e$ and I$_l$) separated by $\textrm{t}_{s}$ and with a relative phase difference $\phi$. The pair of pulses are stored and analyzed directly in the memory. This requires the implementation of partial readouts \cite{Staudt2007a}, achieved by sending two readout pulses ($\pi_2$ and $\pi_3$) separated also by $\textrm{t}_{s}$ (see upper  panel of  Fig. \ref{fig4a}). The resulting output of the memory is  three temporal modes, where the central mode shows coherent interference between the echo of the late pulse from the first read out and the echo of the early pulse induced by the second read out (E${_l}_{1}$ + E${_e}_{2}$).

Figure \ref{fig4b} shows an interference fringe of the central output mode as the relative phase difference $\phi$ is changed for {$\tau = 5\,\mu\mathrm{s}$, $T_S = 12\,\mu\mathrm{s}$, t$_{{s}} = 1\,\mu\mathrm{s}$ and pulse durations of $700\,\mathrm{ns}$}. For each phase, a total of 10 trials are made for statistics. The detection window width t$_v$ is chosen to be $500\,\mathrm{ns}$. The data clearly show coherent behaviour with a weighted sinusoidal fit giving a visibility of $84.0 \pm 5.7\,\%$. 

The coherent nature of the storage is probed further by investigating the fringe visibility as a function of $T_S$. Shown in the lower panel of Fig. \ref{fig4a}, the visibility remains at a constant value of $\bar{\textrm{V}} = 81.7\pm 1.4\,\%$ as $T_S$ is increased, i.e. is independent of the storage efficiency. This behaviour  is typical of ensemble based quantum memories \cite{Staudt2007}, where atomic decoherence decreases the efficiency of recall but not the conditional fidelity. This is true as long as the signal is much greater than the background noise, which may prove challenging to achieve for single photon level inputs. As a comparison, we repeated the experiment while changing the relative phase of the readout pulses, and also obtained an interference fringe with  ${\textrm{V}} = 84\,\%$.

The observed visibilities would result in storage conditional fidelities exceeding $90\,\%$ ($F_C=(1+V)/2$) which, if this could be demonstrated with single photon fields, would be sufficient for applications in quantum communication. The maximal possible visibility should be however $100 \%$. Note that, besides the possible decoherence in the crystal, the visibility is directly affected by the phase noise between the two input time bins and between the two control pulses. To test this phase noise, we performed an interference measurement using two pulse photon echoes (i.e. without transfer to the ground state). The resulting visibilities were similar, confirming that storing in the ground
state does not affect the coherence. Also, this indicates that the main source of decoherence is the limited coherence of the laser itself, and not the storage and retrieval process.

\subsection{Multimode Storage\label{ssec:MM}}

An important feature of a quantum memory is its multimode capacity. A memory exhibiting this property allows for multiplexing which, for example, can increase the success rate of a quantum repeater by a factor given by the multimode capacity \cite{Simon2007a}. Temporal multimodality has been demonstrated with the two level AFC \cite{Usmani2010, Bonarota2011}, but most implementations have used systems that do not have three ground states and are thus not extendible to the full scheme. 

As stated in \cite{Afzelius2010}, the maximum multimode storage capacity (N$_\textrm{max}$) is limited by the number of teeth (N$_\textrm{teeth}$) in the comb. This can be understood in terms of the ratio between the duration of each input mode and the total duration of the pulse train. The former is limited by the comb bandwidth. The latter is limited by the AFC storage time and thus by the width and number of teeth that can fit in the comb bandwidth. 

For the two level AFC experiments, great efforts have been made to increase the comb width \cite{Usmani2010}, up to GHz bandwidth \cite{Saglamyurek2011,Bonarota2011}. However, this is not possible for 3 level AFC experiments, since the comb bandwidth is determined by the hyperfine splitting, which ranges from a few MHz for Pr doped solids to tens of MHz for Eu doped solids. In order to increase the multimode capacity for spin-wave storage, the difficulty is then to reduce as much as possible the AFC peak widths, which requires a narrow line preparation laser. For our system, the limiting hyperfine splitting is the excited state  one ($\sim5\,\mathrm{MHz}$), while our tooth width is fixed by the laser linewidth of  $\sim100\,\mathrm{kHz}$. The ultimate limit on the tooth width is given by the homogenous width of the praseodymium ion $(\sim 2\,\mathrm{kHz})$ \cite{Equall1995}.

Another parameter affecting the storage efficiency that should be taken into account is the temporal duration of the control pulses. In the ideal case, their bandwidth matches that of the comb. In fact, the duration of the $\pi-$pulse, and thus the bandwidth of the memory, is determined by the available $\Omega_R$. For limited $\Omega_R$, efficient transfer between optical and spin-wave atomic excitation requires long control pulses, which in turn requires long input pulses. This limits the number of modes that can be stored. With this in mind, the extension to an efficient multimode spin-wave memory is non-trivial.

The application of several strong input modes to a single class comb can either induce the emission of a two pulse photon echo, accumulated echo, or in the worst case, destroy the comb itself. To avoid the occurrence of these undermining events, the storage of multiple modes is carried out with low power input pulses $(\sim 2\times10^4$ photons per pulse) and the recalled echoes are detected with a single photon detector (SPD, model Count, Laser Components). Neutral density filters $(\mathrm{OD} = 6.5)$ are placed after the crystal to not saturate the detector. Moreover, a double pass AOM is used to temporally gate any noise from the control mode polluting the output mode. For each memory prepared, a total of 500 pulse trains are sent. The electronic signal from the SPD is sent to a time stamping card which collects the arrival times in a histogram.
\begin{figure}[h!]
\begin{center}
\includegraphics[width=0.8\textwidth]{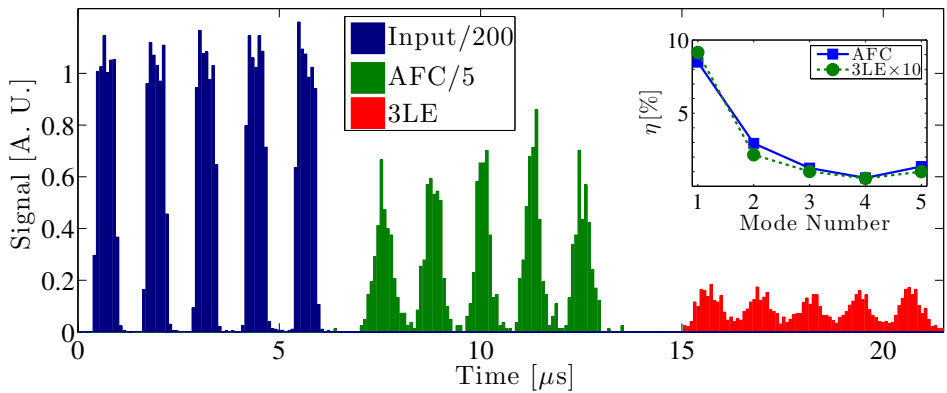}
\caption{Three level storage of 5 temporal modes. The AFC delay is $\tau = 7\mu s$ with $T_S = 7\mu s$. The blue bars from $t = 0 - 6\mu$s shows  5 distinct temporal modes whose height are divided by 200. The green bars show the 2 level AFC echo divided by 5 while the red bars show the 3 level echo. The echoes show clearly the 5 distinct temporal modes. Inset: Three level efficiency vs mode number. For one mode, $\tau = 3\,\mu\mathrm{s}$ and for each increase in temporal mode number, an additional $\mu\mathrm{s}$ is programmed for the AFC storage time.}
\label{fig5}
\end{center}
\end{figure}

Figure 5 shows an example of 5 temporal modes stored and retrieved using the full AFC scheme. An AFC storage time $\tau = 7\,\mu\mathrm{s}$ (corresponding to comb teeth separated by $\Delta = 133\,\mathrm{kHz}$) was programmed, with a bandwidth of $2\,\mathrm{MHz}$ (N$_\textrm{teeth} = 15)$. A total of $1\times10^5\, (2\times10^6)$ pulse trains were used to observe the AFC (3LE) echo. The requirements of a long AFC storage time and of short input pulses contribute to the decrease of the 3LE efficiency with respect to the optimized case of a single mode shown in Fig.~\ref{fig2a}.  Indeed both the two level AFC echo and the transfer efficiency are decreased compared to Fig.~\ref{fig2a}. In fact, the total efficiency for the storage and retrieval of 5 modes is here limited to 0.1$\%$. A confirmation of the fact that the overall efficiency strongly depends on the AFC storage time comes from the inset of Fig. 5, which shows the 3LE efficiency as a function of temporal mode number. As the number of modes is decreased, a shorter AFC storage time is used and higher efficiencies can be achieved. Despite the low efficiency observed, this is to the best of our knowledge the first reported demonstration of spin-wave storage of more than two temporal modes in a doped crystal.

\section{Discussion}
Although spin-wave storage with an echo efficiency of $\eta_\textrm{3LE} = 5.6\,\%$ is, to our knowledge, the highest efficiency reported so far, a significant increase in this efficiency is required to be viable for quantum  communication. To that end, we have strategies  for further efficiency improvement. An increase in optical depth leads to an increase in efficiency. This can be obtained by increasing the physical length of the crystal. Experimentally, preparing a memory with an optically and physically thick crystal is feasible, as demonstrated in \cite{Hedges2010}. An alternate method to increase the optical depth is to increase the doping concentration at the risk of a reduced coherence time due to ion-ion interactions.

The efficiency of the AFC scheme is limited to $54\,\%$ in the  forward direction due to reabsorption of the emitted light. However, in principle, the efficiency can be increased to unity by retrieving the echo in the backward direction \cite{Afzelius2009}. It is realized by the use of counter-propagating control pulses, which for the present set up is readily achievable. 

In this paper, the coherent transfer was mediated by the use of fixed frequency, temporally square transfer $\pi$-pulses. The transfer efficiency is given by the available Rabi frequency which is limited by the available laser intensity. Thus, one option to increase the Rabi frequency is to directly increase the control intensity by increasing the laser power or by using smaller beams. However, as described in \cite{Minar2010}, the transfer efficiency can be optimized with the use of control pulses with a hyperbolic tangent frequency chirp and a hyperbolic secant temporal shape. The benefit is that chirped pulses require less intensity than fixed frequency $\pi$-pulses, for a given transfer efficiency.

An alternate method to increase the efficiency is to place the crystal in an ``impedance matched'' cavity, as proposed in \cite{Afzelius2010a, Moiseev2010} and recently demonstrated in \cite{Sabooni2013}. In principle, the input light can be completely absorbed with a read out efficiency limited only by atomic dephasing.

Our memory time was limited to around 20$\,\mu\mathrm{s}$ due to the inhomogeneous spin broadening.  The storage time can be improved by the use of RF spin echo techniques. Such techniques counteract the effect of dephasing that is introduced by the inhomogeneity of the spin state. Dynamical decoupling techniques have been implemented successfully in {\PrYSO} \cite{Fraval2005, longdell2005} and {Tm$^{3+}$:YAG} \cite{Pascual-Winter2012}. A further increase in storage time can result from the elimination of decoherence sources. For solid state systems, the coherence time can be increased by the application of an external magnetic field of a specific magnitude and angle \cite{Fraval2004, Heinze2011}. In doing so, the first order Zeeman shift becomes vanishingly small, reducing the sensitivity of the Pr$^{3+}$ ions to the host spins. However, this lifts the degeneracy of the hyperfine levels, which could reduce the available bandwidth for the memory. 

We note that, for our experiment, the excited state efficiency $\eta_\textrm{AFC}$ for longer storage times  and hence the multimode capacity will benefit from a narrower band laser. Our laser was limited to producing AFC echoes efficiently $(\sim1\,\%)$ for storage times of up to $7\,\mu\mathrm{s}$.

The next experimental milestone regarding this memory is the storage of single photon level light and nonclassical light into a single spin-wave excitation. The results reported here were obtained with the use of bright input pulses, but in principle can be extended to the quantum light regime as AFC echo memories scale linearly with input power and are in principle noise free \cite{Riedmatten2008}. As well as increasing the efficiency, the major experimental challenge is to filter out pollutive light from the echo mode emitted from other hyperfine levels. This is particularly difficult in {\PrYSO} as the hyperfine levels occur only $10\,\mathrm{MHz}$ away from the signal frequency. For this, narrow band filtering would be required to which a {\PrYSO} crystal is ideally suited, as demonstrated in \cite{Zhang2012, Beavan2012}.

\section{Conclusions}

In conclusion, we have reported on coherent and multi-temporal mode storage of light using the full AFC memory scheme.  Using this scheme, a total efficiency of $\eta_\textrm{3LE} = 5.6\,\%$ was observed, the highest efficiency reported so far. The coherent nature of the spin-wave storage is shown for the first time using a time-bin interference measurement resulting in an average visibility of $81.7\,\%$, independent of the storage time. A total  of 5 temporal modes were stored and recalled on-demand from the memory, the highest number of modes stored as a spin-wave in a doped crystal.  We discussed ways to improve the efficiency and storage time further and the extendibility to the single photon regime. 

\section*{Acknowledgments}
Financial support by the European projects CHIST-ERA QScale and FP7-CIPRIS(MC ITN-287252), as well as by the ERC Starting grant QuLIMA is acknowledged.

\section*{References}
\bibliographystyle{ieeetr}

\end{document}